\documentclass[a4paper]{llncs}

\usepackage{graphicx}
\usepackage{amsmath}
\usepackage{amsfonts}
\usepackage{amssymb}

\newcommand{\SA}{\ensuremath{\mathsf{SA}}}
\newcommand{\LCP}{\ensuremath{\mathsf{LCP}}}
\newcommand{\PLCP}{\ensuremath{\mathsf{PLCP}}}

\title{Sampled Longest Common Prefix Array}

\author{Jouni Sir\'en\thanks{Funded by the Academy of Finland under grant 119815.}}

\institute{Department of Computer Science, University of Helsinki, Finland\\
\email{jltsiren@cs.helsinki.fi}}

\begin{document}

\maketitle

\begin{abstract}
When augmented with the longest common prefix (LCP) array and some other structures, the suffix array can solve many string processing problems in optimal time and space. A compressed representation of the LCP array is also one of the main building blocks in many compressed suffix tree proposals. In this paper, we describe a new compressed LCP representation: the {\em sampled LCP array}. We show that when used with a compressed suffix array (CSA), the sampled LCP array often offers better time/space trade-offs than the existing alternatives. We also show how to construct the compressed representations of the LCP array directly from a CSA.
\end{abstract}

\section{Introduction}

The suffix tree is one of the most important data structures in string processing and bioinformatics. While it solves many problems efficiently, its usefulness is limited by its size: typically 10--20 times the size of the text \cite{Kurtz1999}. Much work has been put on reducing the size, resulting in data structures such as the enhanced suffix array \cite{Abouelhoda2004} and several variants of the compressed suffix tree \cite{Sadakane2007,Russo2008,Fischer2009a,Maekinen2009}.

Most of the proposed solutions are based on three structures: 1) the suffix array, listing the suffixes of the text in lexicographic order; 2) the longest common prefix (LCP) array, listing the lengths of the longest common prefixes of lexicographically adjacent suffixes; and 3) a representation of suffix tree topology. While there exists an extensive literature on compressed suffix arrays (CSA)\footnote{In this paper, we use the term {\em compressed suffix array} to refer to any compressed self-index based on the Burrows-Wheeler transform.} \cite{Navarro2007}, less has been done on compressing the other structures.

Existing proposals to compress the LCP information are based on the permuted LCP (PLCP) array that arranges the entries in text order. While the PLCP array can be compressed, one requires expensive CSA operations to access LCP values through it. In this paper, we describe the sampled LCP array as an alternative to the PLCP-based approaches. Similar to the suffix array samples used in CSAs, the sampled LCP array often offers better time/space trade-offs than the PLCP-based alternatives.

We also modify a recent PLCP construction algorithm \cite{Kaerkkaeinen2009} to work directly with a compressed suffix array. Using it, we can construct any PLCP representation with negligible working space in addition to the CSA and the PLCP. A variant of the algorithm can also be used to construct the sampled LCP array, but requires more working space. While our algorithm is much slower than the alternatives, it is the first LCP construction algorithm that does not require access to the text and the suffix array. This is especially important for large texts, as the suffix array may not be available or the text might not fit into memory.

We begin with basic definitions and background information in Sect.~\ref{sect:background}. Section~\ref{sect:previous} is a summary of previous compressed LCP representations. In Sect.~\ref{sect:irreducible}, we show how to build the PLCP array directly from a CSA. We describe our sampled LCP array in Sect.~\ref{sect:sampled}. Section~\ref{sect:experiments} contains experimental evaluation of our proposals. In Sect.~\ref{sect:comparison}, we compare the sampled LCP array to direct compression of the LCP values. We finish with conclusions and discussion on future work in Sect.~\ref{sect:discussion}.

\section{Background}
\label{sect:background}

A {\em string} $S = S[1,n]$ is a {\em sequence} of {\em characters} from {\em alphabet} $\Sigma = \{ 1, 2, \dots, \sigma \}$. A {\em substring} of $S$ is written as $S[i,j]$. A substring of type $S[1,j]$ is called a {\em prefix}, while a substring of type $S[i,n]$ is called a {\em suffix}. A {\em text} string $T = T[1,n]$ is a string terminated by $T[n] = \$ \not\in \Sigma$ with lexicographic value $0$. The {\em lexicographic order} ''$<$'' among strings is defined in the usual way.

The {\em suffix array (SA)} of text $T[1,n]$ is an array of pointers $\SA[1,n]$ to the suffixes of $T$ in lexicographic order. As an abstract data type, a suffix array is any data structure with similar functionality as the concrete suffix array. This can be defined by an efficient support for the following operations: (a) {\em count} the number of occurrences of a {\em pattern} in the text; (b) {\em locate} these occurrences (or more generally, retrieve a suffix array value); and (c) {\em display} any substring of $T$.

{\em Compressed suffix arrays (CSA)} \cite{Grossi2005,Ferragina2005a} support these operations. Their compression is based on the {\em Burrows-Wheeler transform (BWT)} \cite{Burrows1994}, a permutation of the text related to the SA. The BWT of text $T$ is a sequence $L[1,n]$ such that $L[i] = T[\SA[i] - 1]$, if $\SA[i] > 1$, and $L[i] = T[n] = \$$ otherwise.

The Burrows-Wheeler transform is reversible. The reverse transform is based on a permutation called {\em $LF$-mapping} \cite{Burrows1994,Ferragina2005a}. Let $C[1,\sigma]$ be an array such that $C[c]$ is the number of characters in $\{ \$, 1, 2, \dots, c-1 \}$ occurring in the text. For convenience, we also define $C[0] = 0$ and $C[\sigma + 1] = n$. By using this array and the sequence $L$, we define $LF$-mapping as $LF(i) = C[L[i]] + rank_{L[i]}(L,i)$, where $rank_{c}(L,i)$ is the number of occurrences of character $c$ in prefix $L[1,i]$.

The inverse of $LF$-mapping is $\Psi(i) = select_{c}(L, i - C[c])$, where $c$ is the highest value with $C[c] < i$, and $select_{c}(L,j)$ is the position of the $j$th occurrence of character $c$ in $L$ \cite{Grossi2005}. By its definition, function $\Psi$ is strictly increasing in the range $\Psi_{c} = [C[c] + 1, C[c+1]]$ for every $c \in \Sigma$. Additionally, $T[\SA[i]] = c$ and $L[\Psi(i)] = c$ for every $i \in \Psi_{c}$.

These functions form the backbone of CSAs. As $\SA[LF(i)] = \SA[i]-1$ \cite{Ferragina2005a} and hence $\SA[\Psi(i)] = \SA[i]+1$, we can use these functions to move the suffix array position backward and forward in the sequence. Both of the functions can be efficiently implemented by adding some extra information to a compressed representation of the BWT. Standard techniques \cite{Navarro2007} to support suffix array operations include {\em backward searching} \cite{Ferragina2005a} for {\em count}, and adding a sample of suffix array values for {\em locate} and {\em display}.

Let $lcp(A, B)$ be the length of the longest common prefix of sequences $A$ and $B$. The {\em longest common prefix (LCP) array} of text $T[1,n]$ is the array $\LCP[1,n]$ such that $\LCP[1] = 0$ and $\LCP[i] = lcp(T[\SA[i-1],n], T[\SA[i],n])$ for $i > 1$. The array requires $n \log n$ bits of space, and can be constructed in $O(n)$ time \cite{Kasai2001,Kaerkkaeinen2009}.

\section{Previous Compressed LCP Representations}
\label{sect:previous}

We can exploit the redundancy in LCP values by reordering them in text order. This results in the {\em permuted LCP (PLCP) array}, where $\PLCP[\SA[i]] = \LCP[i]$. The following lemma describes a key property of the PLCP array.

\begin{lemma}[\cite{Kasai2001,Kaerkkaeinen2009}]\label{lemma:plcp}
For every $i \in \{ 2, \dots, n \}$, $\PLCP[i] \ge \PLCP[i-1] - 1$.
\end{lemma}

As the values $\PLCP[i] + 2i$ form a strictly increasing sequence, we can store the array in a bit vector of length $2n$ \cite{Sadakane2007}. Various schemes exist to represent this bit vector in a succinct or compressed form \cite{Sadakane2007,Fischer2009a,Maekinen2009}.

Space-efficiency can also be achieved by sampling every $q$th PLCP value, and deriving the missing values when needed \cite{Khmelev2004}. Assume we have sampled $\PLCP[aq]$ and $\PLCP[(a+1)q]$, and we want to determine $\PLCP[aq+b]$ for some $b < q$. Lemma~\ref{lemma:plcp} states that $\PLCP[aq] - b \le \PLCP[aq+b] \le \PLCP[(a+1)q] + q - b$, so at most $q + \PLCP[(a+1)q] - \PLCP[aq]$ character comparisons are required to determine the missing value. The average number of comparisons over all entries is $O(q)$ \cite{Kaerkkaeinen2009}. By carefully selecting the sampled positions, we can store the samples in $o(n)$ bits, while requiring only $O(\log^{\delta} n)$ comparisons in the worst case for any $0 < \delta \le 1$ \cite{Fischer2009}.

Unfortunately these compressed representations are not very suitable for use with CSAs. The reason is that the LCP values are accessed through suffix array values, and {\em locate} is an expensive operation in CSAs. In addition to that, sampled PLCP arrays require access to the text, using the similarly expensive {\em display}.

Assume that a CSA has SA sample rate $d$, and that it computes $\Psi(\cdot)$ in time $t_{\Psi}$. To retrieve $\SA[i]$, we compute $i, \Psi(i), \Psi^{2}(i), \dots$, until we find a sampled suffix array value. If the sampled value was $\SA[\Psi^{k}(i)] = j$, then $\SA[i] = j - k$. We find a sample in at most $d$ steps, so the time complexity for {\em locate} is $O(d \cdot t_{\Psi})$. Similarly, to retrieve a substring $T[i,i+l]$, we use the samples to get $\SA^{-1}[d \cdot \lfloor \frac{i}{d} \rfloor]$. Then we iterate the function $\Psi$ until we reach text position $i + l$. This takes at most $d + l$ iterations, making the time complexity for {\em display} $O((d + l) \cdot t_{\Psi})$. From these bounds, we get the PLCP access times shown in Table~\ref{table:representations}.\footnote{Some CSAs use $LF$-mapping instead of $\Psi$, but similar results apply to them as well.}

\begin{table}[t!]
\centering
\caption{Time/space trade-offs for (P)LCP representations. $R$ is the number of equal letter runs in BWT, $q$ is the PLCP sample rate, and $0 < \delta \le 1$ is a parameter. The numbers for CSA assume $\Psi$ access time $t_{\Psi}$ and SA sample rate $d$.}
\label{table:representations}
\renewcommand{\tabcolsep}{.1cm}
\begin{tabular}{llll}
\hline\noalign{\smallskip}
 & & \multicolumn{2}{c}{{\bf Access times}} \\
{\bf Representation} & {\bf Space (bits)} & {\bf Using SA} & {\bf Using CSA} \\
\noalign{\smallskip}
\hline
\noalign{\smallskip}
LCP & $n \log n$  & $O(1)$ & $O(1)$ \\
PLCP \cite{Sadakane2007} & $2n + o(n)$  & $O(1)$ & $O(d \cdot t_{\Psi})$ \\
PLCP \cite{Fischer2009a} & $2R \log \frac{n}{R} + O(R) + o(n)$ & $O(1)$ & $O(d \cdot t_{\Psi})$ \\
PLCP \cite{Maekinen2009}  & $2R \log \frac{n}{R} + O(R \log \log \frac{n}{R})$ & $O(\log \log n)$ & $O(d \cdot t_{\Psi} + \log \log n)$ \\
Sampled PLCP \cite{Fischer2009} & $o(n)$  & $O(\log^{\delta} n)$ & $O((d + \log^{\delta} n) \cdot t_{\Psi})$ \\
Sampled PLCP \cite{Khmelev2004} & $\frac{n}{q} \log n$ & $O(q)$ & $O((d + q) \cdot t_{\Psi})$ \\
\noalign{\smallskip}
\hline
\end{tabular}
\end{table}

Depending on the type of index used, $t_{\psi}$ varies from $O(1)$ to $O(\log n)$ in the worst case \cite{Navarro2007}, and is close to 1 microsecond for the fastest indexes in practice \cite{Ferragina2009a,Maekinen2009}. This is significant enough that it makes sense to keep $t_{\Psi}$ in  Table~\ref{table:representations}.

The only (P)LCP representation so far that is especially designed for use with CSAs is Fischer's Wee LCP \cite{Fischer2009} that is basically the $select$ structure from Sadakane's bit vector representation \cite{Sadakane2007}. When the bit vector itself would be required to answer a query, some characters of two lexicographically adjacent suffixes are compared to determine the LCP value. This increases the time complexity, while reducing the size significantly. In this paper, we take the other direction by reducing the access time, while achieving similar compression as in the run-length encoded PLCP variants \cite{Fischer2009a,Maekinen2009}.

\section{Building the PLCP Array from a CSA}
\label{sect:irreducible}

In this section, we adapt the {\em irreducible LCP algorithm} \cite{Kaerkkaeinen2009} to compute the PLCP array directly from a CSA.

\begin{definition}
For $i > 1$, the {\em left match} of suffix $T[\SA[i],n]$ is $T[\SA[i-1],n]$.
\end{definition}

\begin{definition}
Let $T[j,n]$ be the left match of $T[i,n]$. $\PLCP[i]$ is {\em reducible}, if $i, j > 1$ and $T[i-1] = T[j-1]$. If $\PLCP[i]$ is not reducible, then it is {\em irreducible}.
\end{definition}

The following lemma shows why reducible LCP values are called reducible.

\begin{lemma}[\cite{Kaerkkaeinen2009}]\label{lemma:reducible}
If $\PLCP[i]$ is reducible, then $\PLCP[i] = \PLCP[i-1]-1$.
\end{lemma}

The irreducible LCP algorithm works as follows: 1) find the irreducible PLCP values; 2) compute them naively; and 3) fill in the reducible values by using Lemma~\ref{lemma:reducible}. As the sum of the irreducible values is at most $2n \log n$, the algorithm works in $O(n \log n)$ time \cite{Kaerkkaeinen2009}.

The original algorithm uses the text and its suffix array that are expensive to access in a CSA. In the following lemma, we show how to find the irreducible values by using the function $\Psi$ instead.

\begin{lemma}\label{lemma:csa reducible}
Let $T[j,n]$ be the left match of $T[i,n]$. The value $\PLCP[i+1]$ is reducible if and only if $T[i] = T[j]$ and $\Psi(\SA^{-1}[j]) = \Psi(\SA^{-1}[i])-1$.
\end{lemma}

\begin{proof}
Let $x = \SA^{-1}[i]$. Then $x-1 = \SA^{-1}[j]$.

{\em ''If.''} Assume that $T[i] = T[j]$ and $\Psi(x-1) = \Psi(x) - 1$. Then the left match of $T[\SA[\Psi(x)],n] = T[i+1,n]$ is $T[\SA[\Psi(x-1)],n] = T[j+1,n]$. As $i+1 > 1$ and $j+1 > 1$, it follows that $\PLCP[i+1]$ is reducible.

{\em ''Only if.''} Assume that $\PLCP[i+1]$ is reducible, and let $T[k,n]$ be the left match of $T[i+1,n]$. Then $k > 1$ and $T[k-1] = T[i]$. As $T[k-1,n]$ and $T[i,n]$ begin with the same character, and $T[k,n]$ is the left match of $T[i+1,n]$, there cannot be any suffix $S$ such that $T[k-1,n] < S < T[i,n]$. But now $j = k-1$, and hence $T[i] = T[j]$. Additionally,
\begin{displaymath}
\Psi(\SA^{-1}[j]) = \Psi(\SA^{-1}[k-1]) = \SA^{-1}[k] = \SA^{-1}[i+1] - 1 = \Psi(\SA^{-1}[i]) - 1.
\end{displaymath}
The lemma follows. $\Box$
\end{proof}

The algorithm is given in Fig.~\ref{fig:irreducible}. We maintain invariant $x = \SA^{-1}[i]$, and scan through the CSA in text order. If the conditions of Lemma~\ref{lemma:csa reducible} do not hold for $T[i,n]$, then $\PLCP[i+1]$ is irreducible, and we have to compute it. Otherwise we reduce $\PLCP[i+1]$ to $\PLCP[i]$. To compute an irreducible value, we iterate $(\Psi^{k}(b-1), \Psi^{k}(b))$ for $k = 0, 1, 2, \dots$, until $T[\Psi^{k}(b-1)] \ne T[\Psi^{k}(b)]$. When this happens, we return $k$ as the requested LCP value. As we compute $\Psi(\cdot)$ for a total of $O(n \log n)$ times, we get the following theorem.

\begin{figure}[t!]
\begin{minipage}[t]{.5\textwidth}
\begin{tabbing}
mm\=mn\=mn\= \kill
--- Compute the PLCP array \\
1 \> $\PLCP[1] \leftarrow 0$ \\
2 \> $(i, x) \leftarrow (1, \SA^{-1}[1])$ \\
3 \> {\bf while} $i < n$ \\
4 \> \> $\Psi_{c} \leftarrow$ rangeContaining($x$) \\
5 \> \> {\bf if} $x-1 \not\in \Psi_{c}$ {\bf or} $\Psi(x-1) \ne \Psi(x) - 1$ \\
6 \> \> \> $\PLCP[i+1] \leftarrow$ lcp($\Psi(x)$) \\
7 \> \> {\bf else} $\PLCP[i+1] \leftarrow \PLCP[i]-1$ \\
8 \> \> $(i, x) \leftarrow (i+1, \Psi(x))$ \\
\end{tabbing}
\end{minipage}
\hspace{5pt}
\begin{minipage}[t]{.5\textwidth}
\begin{tabbing}
mm\=mn\=mn\= \kill
--- Compute an LCP value \\
9 \> {\bf def} lcp($b$) \\
10 \> \> $(a, k) \leftarrow (b-1, 0)$ \\
11 \> \> $\Psi_{c} \leftarrow$ rangeContaining($b$) \\
12 \> \> {\bf while} $a \in \Psi_{c}$ \\
13 \> \> \> $(a, b, k) \leftarrow (\Psi(a), \Psi(b), k+1)$ \\
14 \> \> \> $\Psi_{c} \leftarrow$ rangeContaining($b$) \\
15 \> \> {\bf return} $k$ \\
\end{tabbing}
\end{minipage}

\caption{The irreducible LCP algorithm for using a CSA to compute the PLCP array. Function rangeContaining($x$) returns $\Psi_{c} = [C[c]+1, C[c+1]]$ where $x \in \Psi_{c}$.}\label{fig:irreducible}
\end{figure}

\begin{theorem}
Given a compressed suffix array for a text of length $n$, the irreducible LCP algorithm computes the PLCP array in $O(n \log n \cdot t_{\Psi})$ time, where $t_{\Psi}$ is the time required for accessing $\Psi$. The algorithm requires $O(\log n)$ bits of working space in addition to the CSA and the PLCP array.
\end{theorem}

We can use the algorithm to build any PLCP representation from Table~\ref{table:representations} directly. The time bound is asymptotically tight, as shown in the following lemma.

\begin{lemma}[Direct extension of Lemma 5 in \cite{Kaerkkaeinen2009}]
For an order-$k$ {\em de Bruijn sequence} on an alphabet of size $\sigma$, the sum of all irreducible PLCP values is $n (1 - 1/\sigma) \log_{\sigma} n - O(n)$.
\end{lemma}

The sum of irreducible PLCP values of a random sequence should also be close to $n (1 - 1/\sigma) \log_{\sigma} n$. The probability that the characters preceding a suffix and its left match differ, making the PLCP value irreducible, is $(1 - 1/\sigma)$. On the other hand, the average irreducible value should be close to $\log_{\sigma} n$ \cite{Fayolle2005}. For a text generated by an order-$k$ Markov source with $H$ bits of entropy, the estimate becomes $n (1 - 1/\sigma') (\log n) / H$. Here $\sigma'$ is the effective alphabet size, defined by the probability $1 / \sigma'$ that two characters sharing an order-$k$ context are identical.

The following proposition shows that large-scale repetitiveness reduces the sum of the irreducible values, and hence improves the algorithm performance.

\begin{proposition}\label{prop:repetitive}
For a concatenation of $r$ copies of text $T[1,n]$, the sum of irreducible PLCP values is $s + (r-1)n$, where $s$ is the sum of the irreducible PLCP values of $T$.
\end{proposition}

\begin{proof}
Let $\mathcal{T} = T_{1} T_{2} \cdots T_{r}$ be the concatenation, $\mathcal{T}_{a,i}$ the suffix starting at $T_{a}[i]$, and $\PLCP_{a}[i]$ the corresponding PLCP value. Assume that $T_{r}[n]$ is lexicographically greater than the other end markers, but otherwise identical to them.

For every $i$, the suffix array of $\mathcal{T}$ contains a range with values $\mathcal{T}_{1,i}, \mathcal{T}_{2,i}, \dots, \mathcal{T}_{r,i}$ \cite{Maekinen2009}. Hence for any $a > 1$ and any $i$, the left match of $\mathcal{T}_{a,i}$ is $\mathcal{T}_{a-1,i}$, making the PLCP values reducible for almost all of the suffixes of $T_{2}$ to $T_{r}$. The exception is that $\mathcal{T}_{2,1}$ is irreducible, as its left match is $\mathcal{T}_{1,1}$, and hence $\PLCP_{2}[1] = (r-1)n$.

Let $T[j,n]$ be the left match of $T[i,n]$ in the suffix array of $T$. Then the left match of $\mathcal{T}_{1,i}$ is $\mathcal{T}_{r,j}$, and $\PLCP_{1}[i] = \PLCP[i]$. Hence the sum of the irreducible values corresponding to the suffixes of $T_{1}$ is $s$. $\Box$
\end{proof}

\section{Sampled LCP Array}
\label{sect:sampled}

By Lemmas \ref{lemma:plcp} and \ref{lemma:reducible}, the local maxima in the PLCP array are among the irreducible values, and the local minima are immediately before them.

\begin{definition}
The value $\PLCP[i]$ is {\em maximal}, if it is irreducible, and {\em minimal}, if either $i = n$ or $\PLCP[i+1]$ is maximal.
\end{definition}

\begin{lemma}\label{lemma:minimal reducible}
If $\PLCP[i]$ is non-minimal, then $\PLCP[i] = \PLCP[i+1] + 1$.
\end{lemma}

\begin{proof}
If $\PLCP[i]$ is non-minimal, then $\PLCP[i+1]$ is reducible. The result follows from Lemma~\ref{lemma:reducible}. $\Box$
\end{proof}

In the following, $R$ is the number of equal letter runs in BWT.

\begin{lemma}\label{lemma:minimal sum}
The number of minimal PLCP values is $R$.
\end{lemma}

\begin{proof}
Lemma~\ref{lemma:csa reducible} essentially states that $\PLCP[i+1]$ is reducible, if and only if $L[\Psi(\SA^{-1}[i])] = T[i] = T[j] = L[\Psi(\SA^{-1}[j])] = L[\Psi(\SA^{-1}[i]) - 1]$, where $T[j,n]$ is the left match of $T[i,n]$. As this is true for $n - R$ positions $i$, there are exactly $R$ irreducible values. As every maximal PLCP value can be reduced to the next minimal value, and vice versa, the lemma follows. $\Box$
\end{proof}

\begin{lemma}
The sum of minimal PLCP values is $S - (n - R)$, where $S$ is the sum of maximal values.
\end{lemma}

\begin{proof}
From Lemmas \ref{lemma:minimal reducible} and \ref{lemma:minimal sum}. $\Box$
\end{proof}

If we store the minimal PLCP values in SA order, and mark their positions in a bit vector, we can use them in a similar way as the SA samples. If we need $\LCP[i]$, and $\LCP[\Psi^{k}(i)]$ is a sampled position for the smallest $k \ge 0$, then $\LCP[i] = \LCP[\Psi^{k}(i)] + k$. As $k$ can be $\Theta(n)$ in the worst case, the time bound is $O(n \cdot t_{\Psi})$.

To improve the performance, we sample one out of $d' = n / R^{1-\varepsilon}$ consecutive non-minimal values for some $\varepsilon > 0$. Then there are $R$ minimal samples and at most $R^{1-\varepsilon}$ extra samples. We mark the sampled positions in a bit vector of Raman et al.~\cite{Raman2002}, taking at most $(1 + o(1)) \cdot R \log \frac{n}{R} + O(R) + o(n)$ bits of space. Checking whether an LCP entry has been sampled takes $O(1)$ time.

We use $\delta$ codes \cite{Elias1975} to encode the actual samples. As the sum of the minimal values is at most $2 n \log n$, these samples take at most
\begin{displaymath}
R \log \frac{2 n \log n}{R} + O\left( R \log \log \frac{n}{R} \right) \le R \log \frac{n}{R} + O(R \log \log n)
\end{displaymath}
bits of space. The extra samples require at most $\log n + O(\log \log n)$ bits each. To provide fast access to the samples, we can use dense sampling \cite{Ferragina2007} or directly addressable codes \cite{Brisaboa2009}. This increases the size by a factor of $1 + o(1)$, making the total for samples $(1 + o(1)) \cdot R \log \frac{n}{R} + O(R \log \log n) + o(R \log n)$ bits of space.

We find a sampled position in at most $n / R^{1-\varepsilon}$ steps. By combining the size bounds, we get the following theorem.

\begin{theorem}\label{theorem:sampled}
Given a text of length $n$ and a parameter $0 < \varepsilon < 1$, the sampled LCP array requires at most $(2 + o(1)) \cdot R \log \frac{n}{R} + O(R \log \log n) + o(R \log n) + o(n)$ bits of space, where $R$ is the number of equal letter runs in the BWT of the text. When used with a compressed suffix array, retrieving an LCP value takes at most $O((n / R^{1-\varepsilon}) \cdot t_{\Psi})$ time, where $t_{\Psi}$ is the time required for accessing $\Psi$.
\end{theorem}

By using the BSD representation \cite{Gupta2007} for the bit vector, we can remove the $o(n)$ term from the size bound with a slight loss of performance.

When the space is limited, we can afford to sample the LCP array denser than the SA, as SA samples are larger than LCP samples. In addition to the mark in the bit vector, an SA sample requires $2 \log \frac{n}{d}$ bits of space, while an LCP sample takes just $\log v + O(\log \log v)$ bits, where $v$ is the sampled value.

The LCP array can be sampled by a two-pass version of the irreducible LCP algorithm. On the first pass, we scan the CSA in suffix array order to find the minimal samples. Position $x$ is minimal, if $x$ is the smallest value in the corresponding $\Psi_{c}$, or if $\Psi(x-1) \ne \Psi(x) - 1$. As we compress the samples immediately, we only need $O(\log n)$ bits of working space. On the second pass, we scan the CSA in text order, and store the extra samples in an array. Then we sort the array to SA order, and merge it with the minimal samples. As the number of extra samples is $o(R)$, we need $o(R \log n)$ bits of working space.

\begin{theorem}\label{theorem:slcp construction}
Given a compressed suffix array for a text of length $n$, the modified irreducible LCP algorithm computes the sampled LCP array in $O(n \log n \cdot t_{\Psi})$ time, where $t_{\Psi}$ is the time required for accessing $\Psi$. The algorithm requires $o(R \log n)$ bits of working space in addition to the CSA and the samples, where $R$ is the number of equal letter runs in the BWT of the text.
\end{theorem}

\section{Implementation and Experiments}
\label{sect:experiments}

We have implemented the sampled LCP array, a run-length encoded PLCP array, and their construction algorithms as a part of the RLCSA \cite{Siren2009}.\footnote{The implementation is available at \url{http://www.cs.helsinki.fi/group/suds/rlcsa/}.} For PLCP, we used the same run-length encoded bit vector as in the RLCSA. For the sampled LCP, we used a gap encoded bit vector to mark the sampled positions, and a stripped-down version of the same vector for storing the samples.

To avoid redundant work, we compute minimal instead of maximal PLCP values, and interleave their computation with the main loop. To save space, we only use {\em strictly minimal} PLCP values with $\PLCP[i] < \PLCP[i+1] + 1$ as the minimal samples. When sampling the LCP array, we make both of the passes in text order, and store all the samples in an array before compressing them.

For testing, we used a 2.66 GHz Intel Core 2 Duo E6750 system with 4 GB of memory (3.2 GB visible to OS) running a Fedora-based Linux with kernel 2.6.27. The implementation was written in C++, and compiled on g++ version 4.1.2. We used four data sets: human DNA sequences ({\em dna}) and English language texts ({\em english}) from the Pizza \& Chili Corpus \cite{Ferragina2009a}, the Finnish language Wikipedia with version history ({\em fiwiki}) \cite{Siren2009}, and the genomes of 36 strains of Saccharomyces paradoxus ({\em yeast}) \cite{Maekinen2009}.\footnote{The yeast genomes were obtained from the Durbin Research Group at the Sanger Institute (\url{http://www.sanger.ac.uk/Teams/Team71/durbin/sgrp/}).} When the data set was much larger than 400 megabytes, a 400 MB prefix was used instead. Further information on the data sets can be found in Table~\ref{table:data}.

Only on the {\em dna} data set, the sum of the minimal values was close to the entropy-based estimate.
On the highly repetitive {\em fiwiki} and {\em yeast} data sets, the difference between the estimate and the measurement was very large, as predicted by Proposition~\ref{prop:repetitive}. Even regular English language texts contained enough large-scale repetitiveness that the sum of the minimal values could not be adequately explained by the entropy of the texts. This suggests that, for many real-world texts, the number of runs in BWT is a better compressibility measure than the empirical entropy.

\begin{table}[p]
\centering
\caption{Properties of the data sets. $H_{5}$ is the order-$5$ empirical entropy, $\sigma'$ the corresponding effective alphabet size, $\#$ the number of (strictly) minimal values, and $S$ the sum of those values. $S' = n (1 - 1/\sigma') (\log n) / H_{5} - n / \sigma'$ is an entropy-based estimate for the sum of the minimal values. The construction times are in seconds.}
\label{table:data}
\renewcommand{\tabcolsep}{.1cm}
\begin{tabular}{lrrrrrrrrrrrrr}
\hline\noalign{\smallskip}
 & & & \multicolumn{3}{c}{{\bf Estimates}} & & \multicolumn{3}{c}{{\bf Minimal values}} & & \multicolumn{3}{c}{{\bf Strictly minimal}} \\
{\bf Name} & {\bf MB} & &
$\mathbf{H_{5}}$       & $\mathbf{\sigma'}$    & $\mathbf{S' / 10^{6}}$ & &
$\mathbf{\# / 10^{6}}$ & $\mathbf{S / 10^{6}}$ & $\mathbf{S / n}$ & &
$\mathbf{\# / 10^{6}}$ & $\mathbf{S / 10^{6}}$ & $\mathbf{S / n}$ \\
\noalign{\smallskip}
\hline
\noalign{\smallskip}
english & 400 & & 1.86 & 2.09 & 3167 & & 156.35 & 1736 & 4.14 & &  99.26 & 1052 & 2.51\\
fiwiki  & 400 & & 1.09 & 1.52 & 3490 & &   1.79 &  273 & 0.65 & &   1.17 &  117 & 0.28\\
dna     & 385 & & 1.90 & 3.55 & 4252 & & 243.49 & 3469 & 8.59 & & 158.55 & 2215 & 5.48\\
yeast   & 409 & & 1.87 & 3.34 & 4493 & &  15.64 &  520 & 1.21 & &  10.05 &  299 & 0.70\\
\noalign{\smallskip}
\hline
\end{tabular}
\vspace{8pt}
\renewcommand{\tabcolsep}{.1cm}
\begin{tabular}{lrllrrrrrr}
\hline\noalign{\smallskip}
 & & \multicolumn{2}{c}{{\bf Sample rates}} & & \multicolumn{2}{c}{{\bf PLCP}} & & \multicolumn{2}{c}{{\bf Sampled LCP}} \\
{\bf Name} & & {\bf SA} & {\bf LCP} & & {\bf Time} & {\bf MB/s} & & {\bf Time} & {\bf MB/s} \\
\noalign{\smallskip}
\hline
\noalign{\smallskip}
english & & 8, 16, 32, 64     & 8, 16       & & 1688 & 0.24 & & 2104 & 0.19 \\
fiwiki  & & 64, 128, 256, 512 & 32, 64, 128 & &  327 & 1.22 & &  533 & 0.75 \\
dna     & & 8, 16, 32, 64     & 8, 16       & & 3475 & 0.11 & & 3947 & 0.10 \\
yeast   & & 32, 64, 128, 256  & 16, 32, 64  & &  576 & 0.71 & &  890 & 0.46 \\
\noalign{\smallskip}
\hline
\end{tabular}
\end{table}

\begin{figure}[p]
\includegraphics[width=0.49\textwidth]{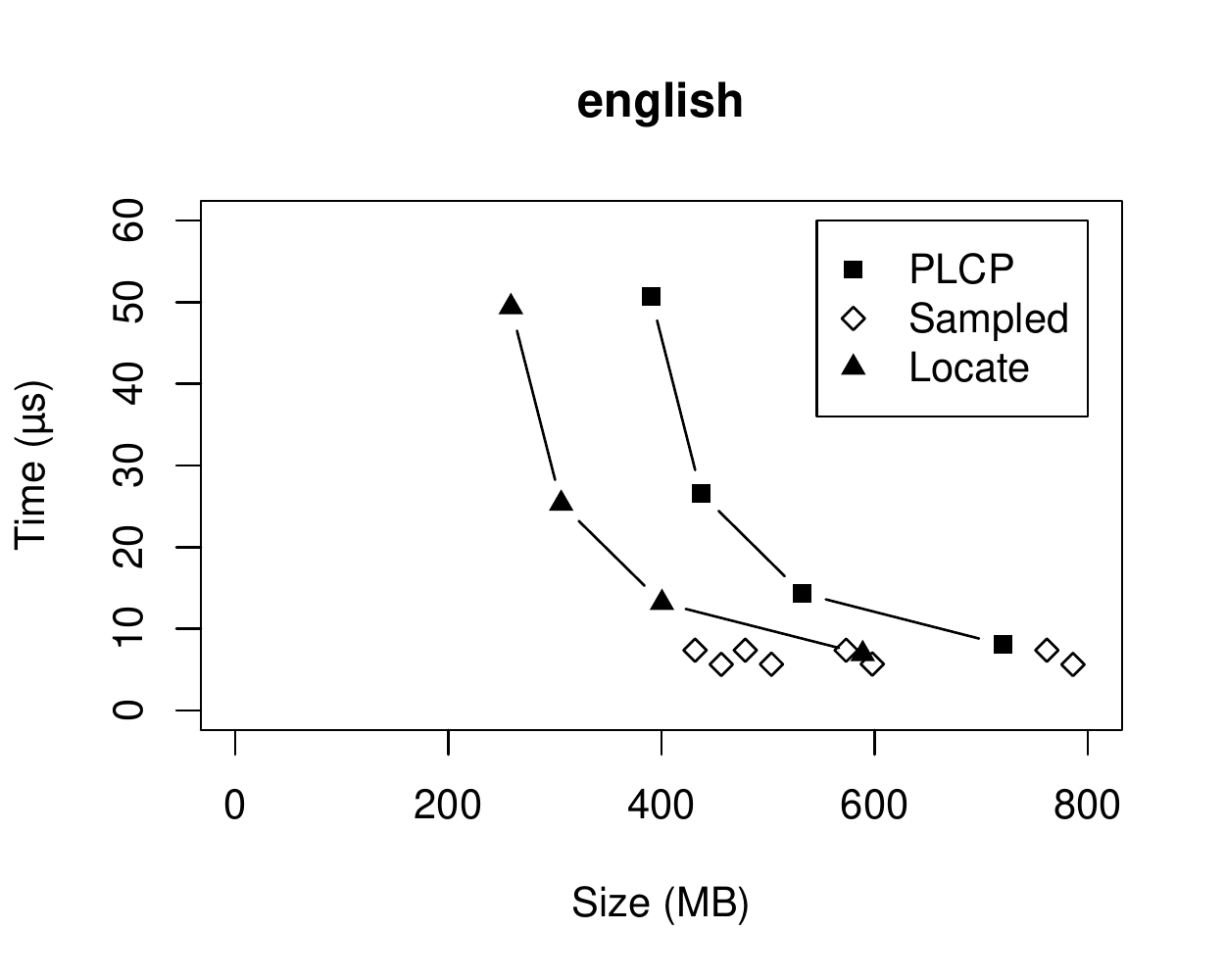}
\includegraphics[width=0.49\textwidth]{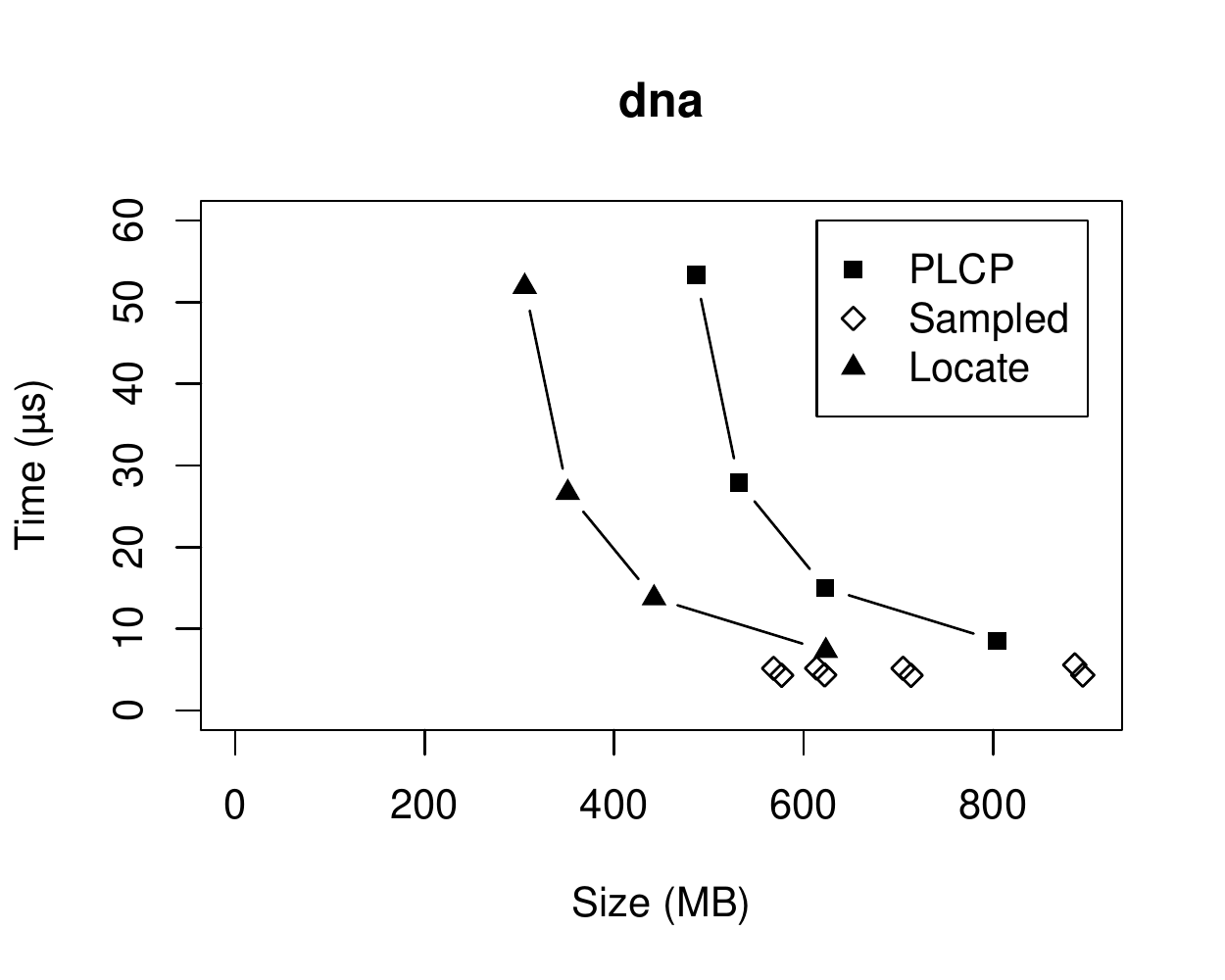}

\includegraphics[width=0.49\textwidth]{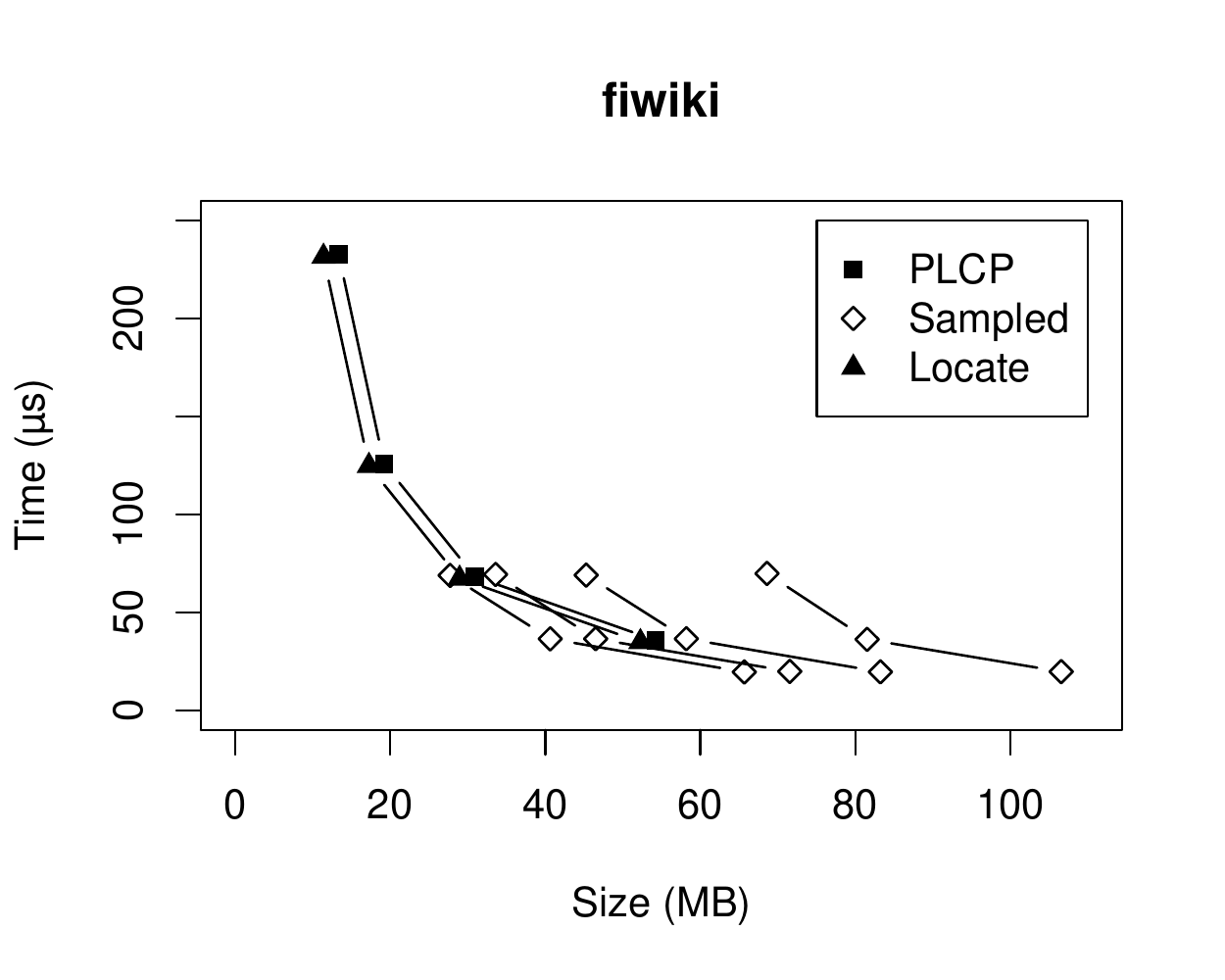}
\includegraphics[width=0.49\textwidth]{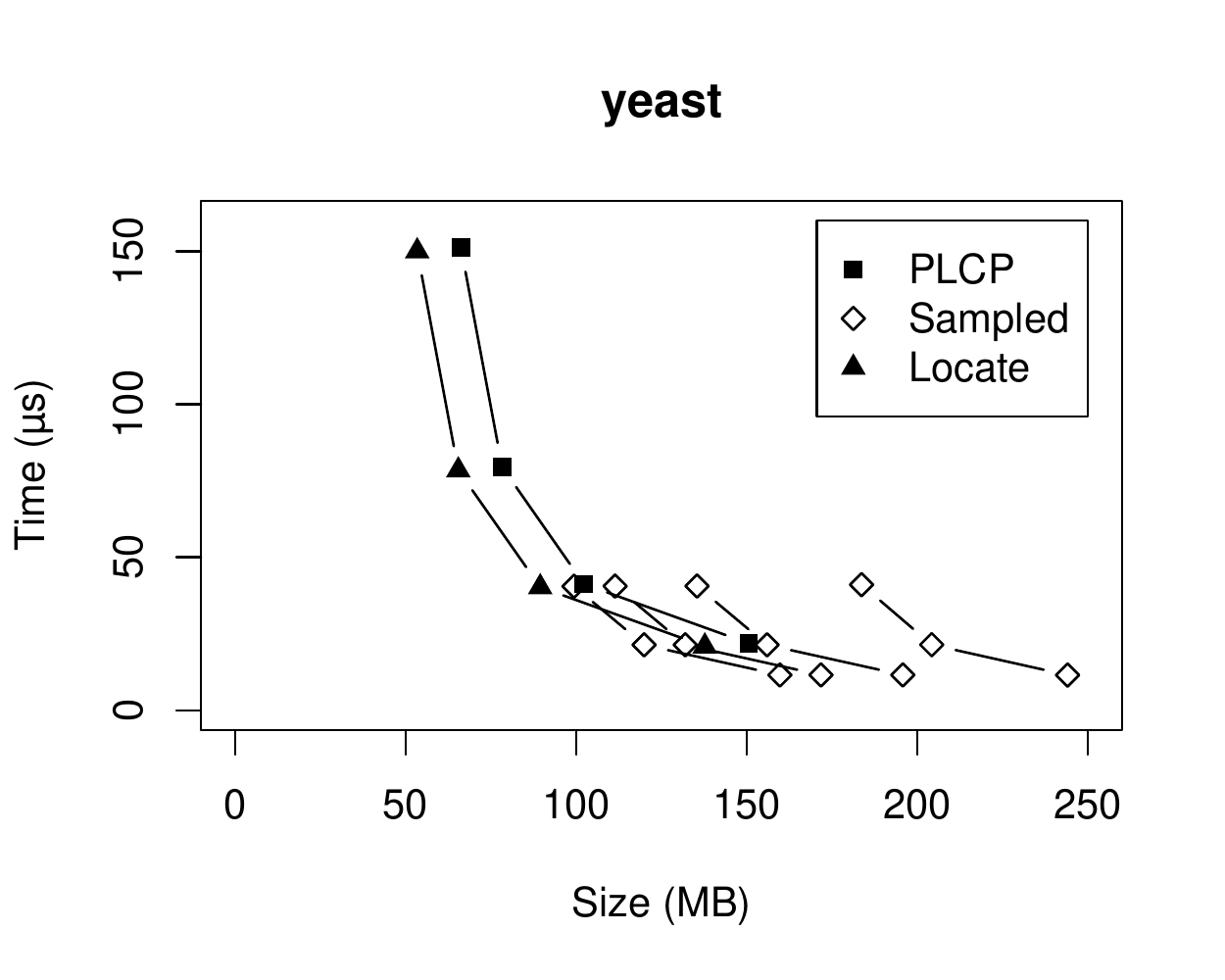}

\caption{Time/space trade-offs for retrieving an LCP or SA value. The times are averages over $10^{6}$ random queries. Sampled LCP results are grouped by SA sample rate.}\label{fig:experiments}
\end{figure}

The sum of minimal PLCP values was a good estimate for PLCP construction time. LCP sampling was somewhat slower because of the second pass. Both algorithms performed reasonably well on the highly repetitive data sets, but were much slower on the regular ones. The overall performance was roughly an order of magnitude worse than for the algorithms using plain text and SA \cite{Kaerkkaeinen2009}.

We measured the performance of the sampled LCP array and the run-length encoded PLCP array on each of the data sets. We also measured the {\em locate} performance of the RLCSA to get a lower bound for the time and space of any PLCP-based approach. The results can be seen in Fig.~\ref{fig:experiments}.

The sampled LCP array outperformed PLCP on {\em english} and {\em dna}, where most of the queries were resolved through minimal samples. On {\em fiwiki} and {\em yeast}, the situation was reversed. As many extra samples were required to get reasonable performance, increasing the size significantly, the sampled LCP array had worse time/space trade-offs than the PLCP array.

While we used RLCSA in the experiments, the results generalize to other types of CSA as well. The reason for this is that, in both PLCP and sampled LCP, the time required for retrieving an LCP value depends mostly on the number of iterations of $\Psi$ required to find a sampled position.

\section{Comparison with Direct LCP Compression}
\label{sect:comparison}

In a recent proposal \cite{Canovas2010}, the entire LCP array was compressed by using directly addressable codes (DAC-LCP) \cite{Brisaboa2009}. The resulting structure was much faster than the other compressed LCP representations, requiring less than a microsecond to access an LCP value. On the other hand, DAC-LCP was also much larger: 6 to 8 bits per character.

To compare the sampled LCP array to DAC-LCP, we downloaded the same data sets as DAC-LCP was tested on. This included 100 MB prefixes of XML data ({\em xml}), human DNA and protein sequences ({\em dna} and {\em proteins}) and source code ({\em sources}) from Pizza \& Chili Corpus. We then sampled the LCP array with sample rate $16$ on each of the data sets. DAC-LCP sizes were visually estimated from the reported results \cite{Canovas2010}. The results are in Table~\ref{table:comparison}.

\begin{table}[t]
\centering
\caption{The sizes of DAC-LCP and two versions of the sampled LCP array in bits per character on 100 MB data sets. $\#$ is the total number of samples.}
\label{table:comparison}
\renewcommand{\tabcolsep}{.1cm}
\begin{tabular}{lrrrr}
\hline\noalign{\smallskip}
{\bf Name} & $\mathbf{\# / n}$ & {\bf Sampled} & {\bf Sampled 2} & {\bf DAC-LCP} \\
\noalign{\smallskip}
\hline
\noalign{\smallskip}
dna        &              0.41 &          5.48 &            3.44 &           5.8 \\
proteins   &              0.44 &          4.39 &            4.16 &           7.0 \\
sources    &              0.17 &          2.33 &            2.33 &           7.5 \\
xml        &              0.13 &          1.85 &            2.09 &           7.8 \\
\noalign{\smallskip}
\hline
\end{tabular}
\end{table}

In addition to the implemented version of the sampled LCP array (Sampled), we also estimated the size of another variant (Sampled 2). Instead of a gap-encoded bit vector to mark the sampled positions, we use the rank/select implementation of González \cite{Gonzalez2005} on a plain bit vector. This takes a total of $1.05n$ bits for a text of length $n$. To store the samples, we use directly addressable codes, estimating the average size of an LCP value to be the same as in DAC-LCP. These results can also be found in Table~\ref{table:comparison}.

On each of the data sets, the sampled LCP array was clearly smaller than DAC-LCP. The difference was smaller on the mostly random {\em dna} and {\em proteins} data sets than on the more structured {\em sources} and {\em xml} data sets. The reason is that on random data, LCP values are small and the number of samples is large, decreasing the size of DAC-LCP and increasing the size of the sampled LCP array, respectively. This also suggests that DAC-LCP becomes very large on highly repetitive data sets, where most of the LCP values are large.

\section{Discussion}
\label{sect:discussion}

We have described the sampled LCP array, and shown that it offers better time/space trade-offs than the PLCP-based alternatives, when the number of extra samples required for dense sampling is small. Based on the experiments, it seems that one should use the sampled LCP array for regular texts, and a PLCP-based representation for highly repetitive texts. DAC-LCP is also a good choice for regular texts, if performance is more important than compression.

We have also shown that it is feasible to construct the (P)LCP array directly from a CSA. While the earlier algorithms are much faster, it is now possible to construct the (P)LCP array for larger texts than before, and the performance is still comparable to that of direct CSA construction \cite{Siren2009}. On a multi-core system, it is also easy to get extra speed by parallelizing the construction.

It is possible to maintain the (P)LCP array when merging two CSAs. The important observation is that an LCP value can only change, if the left match changes in the merge. An open question is, how much faster the merging is, both in the worst case and in practice, than rebuilding the (P)LCP array.

While the suffix array and the LCP array can be compressed to a space relative to the number of of equal letter runs in BWT, no such representation is known for suffix tree topology. This is the main remaining obstacle in the way to compressed suffix trees optimized for highly repetitive texts.

\bibliographystyle{plain}
\bibliography{paper}

\end{document}